`relaced' because original file corrupted in transit

\documentstyle[11pt]{article}
\def\ns{\normalsize}
\begin{document}
\begin{center}
\LARGE
{\bf The End for Extended Inflation?} \\
\vspace{.5cm}
\normalsize
\large{ANDREW R.~LIDDLE$^a$ AND DAVID H.~LYTH$^b$} \\
\normalsize
\vspace{.15cm}
{\em $^a$School of Mathematical and Physical Sciences, \\
University of Sussex, \\ Brighton BN1 9QH.~~~~U.~K.}\\
\vspace{.3cm}
{\em $^b$School of Physics and Materials,\\University of Lancaster,\\
Lancaster LA1 4YB.~~~~U.~K.}
\end{center}
\vspace{0.5 cm}
\begin{center}
{\bf INTRODUCTION}
\end{center}

\vspace{12pt}
\vspace{-.125in}

The most dramatic addition to the inflationary zoo in recent years has been
the extended inflation model$^{1,2}$. At a time when viable models were
dominated by the notion of scalar fields slow-rolling in almost flat
potentials, extended inflation reintroduced the concept of inflation occurring
via a first order phase transition. In such models, inflation is driven by a
scalar field which gets `hung up' in a metastable vacuum state of energy
density $M^4$, with inflation ending via the nucleation of bubbles of true
vacuum, via quantum tunnelling, which can later grow and collide to reheat the
universe into an isotropic Friedmann stage.

There are several reasons why this can be an interesting thing to achieve.
First of all, the possibility arises that inflation can be reintroduced to
realistic particle physics by identifying the inflaton field as a GUT Higgs
particle or equivalent. Secondly, one has obviated the need for a flat
potential (usually a requirement from a combination of sufficient inflation
and small density fluctuations), and so might hope to avoid the fine-tuning
problems that inflation is so commonly accused of possessing$^3$. And thirdly,
there was the hope (which alas has not proven realisable$^4$), that
thermalisation might involve sufficient numbers of astrophysically large
bubbles as to contribute to the `bubbly' structure seen in galaxy redshift
surveys.

The key ingredient of extended inflation is the introduction of a means which
allows the first order phase transition to complete in an inflationary
universe. It has long been known that this is hard to achieve in general
relativity, where Guth's original model$^5$ was implemented, due to a conflict
between requiring a low enough bubble nucleation rate to obtain sufficient
inflation and a high enough one to enable the bubbles to meet and bring
inflation to an end. This is most effectively characterised by defining a
nucleation parameter $\epsilon = \Gamma/H^4$ where $\Gamma$ is the nucleation
rate per unit volume per unit time (a constant barring very unusual and
hard-to-realise field interactions$^6$) and $H$ is the Hubble parameter. Hence
$\epsilon$ is the number of bubbles nucleated per Hubble volume per Hubble
time. One has competing requirements
\begin{tabbing}
\hspace*{3cm} \= {\em Sufficient inflation \hspace*{2cm}} \= $\Longrightarrow$
	\= $\epsilon \ll 1$\\
\> {\em Completion of phase transition} \> $\Longrightarrow$ \= $\epsilon
	\simeq {\cal O}(1)$
\end{tabbing}
In general relativity, one has the unfortunate circumstance that both $\Gamma$
and $H$ are constant, and so these requirements cannot both be satisfied.

What La and Steinhardt realised$^1$ was that one requires a theory in which
$H$ varies during inflation. This can be achieved by going to an `extended'
gravitational theory. Their original model was based on the
Jordan--Brans--Dicke (JBD) theory, which gives $a \propto t^{\omega + 1/2}$
and hence $\epsilon \propto t^4$. With $\epsilon$ growing with time, one can
arrange an early phase in which bubble nucleation is suppressed, allowing
sufficient inflation to occur before $\epsilon$ grows to unity to bring the
phase transition to an end. Note that although $\epsilon$ increases quite
rapidly with time, its increase with comoving scale is rather weak, especially
for large $\omega$.

{\bf Definition}: In this article, `extended inflation' normally refers to all
of a wide class of models which modify gravity to permit a first order phase
transition to an end. Often models with dynamics mimicking the JBD theory will
be discussed illustratively, and occasionally cold dark matter (CDM) will be
chosen as an illustrative choice of matter content for the universe.

\vspace{12pt}
\begin{center}
{\bf CONSTRAINTS ON EXTENDED INFLATION}
\end{center}

\vspace{12pt}
\vspace{-0.16in}
Extended inflation is subject to two main constraints.
\begin{itemize}
\item Bubbles nucleated early in inflation are caught up in the subsequent
expansion and swept up to astrophysical sizes. The bubble distribution is
constrained by the isotropy of the microwave background.
\item As in any inflation model, there are density perturbations generated by
quantum fluctuations, in this case predominantly in the Brans--Dicke field. In
extended inflation the predicted spectrum is `tilted' from flat, and there are
also significant long wavelength gravitational waves generated. These two
perturbation spectra lead to constraints both from the microwave background
and from large scale structure data.
\end{itemize}

Before proceeding onto an up-to-date account of these constraints, it is
remarking on the classes of extended inflation model which exist. Soon after
the original model was devised, the bubble spectrum was derived and subjected
to a heuristic constraint$^{7,8}$ that no more than $10^{-4}$ of the universe
by volume should end up in bubbles larger than the horizon size at decoupling.
Although an incredibly conservative constraint, with no particular reference
to any experiment, this produced the strong limit $\omega \le 30 + \log_{10}
M/m_{Pl}$ where $M$ is, once again, the inflaton mass scale. Combined with
present day solar system experiments$^9$ requiring $\omega$ in excess of 500,
the original model is clearly excluded.

However, model building proved fairly easy, because the conflict originates in
constraints applied at wildly differing times --- the bubble constraint during
inflation at perhaps $10^{-30}$ sec and the solar system limit at the present
time of $10^{17}$ sec. With plenty of room for manoeuvre, two strategies
developed.
\begin{enumerate}
\item $\omega$ really is a constant less than say 25, but is concealed from
present day observation by some mechanism, {\it eg}
	\begin{enumerate}
	\item A potential for $\Phi$.~$^8$
	\item Altered couplings to the invisible sector.$^{10}$
	\item A scale-invariant theory.$^{11}$
	\end{enumerate}
\item $\omega$ is not a constant,
	\begin{enumerate}
	\item Barrow--Maeda $\omega(\Phi)$ model.$^{12}$
	\item Steinhardt--Accetta hyperextended inflation.$^{13}$
	\end{enumerate}
\end{enumerate}

During inflation the first class act exactly as JBD theories, and produce the
same void spectrum. The second class produce a spectrum that requires
case-by-case examination, usually numerically$^{14}$.

The main moral of this article is that improved large scale structure and
microwave background anisotropy measurements, especially the advent of COBE,
have markedly changed this situation, and there are now strong constraints
from structure formation which act in the opposite direction to the bubble
constraints. Because the constraints act at the same time, much of the room to
manoeuvre in model building has been eliminated and many models are now
excluded$^{15}$.

A heuristic way to see the potential conflict is to realise that in the GR
limit both bubble and perturbation spectra are scale-invariant. The same
processes invoked to suppress the bubbles growth from scale-invariance will
necessarily break scale-invariance of the spectra, removing short-scale power.
With the COBE measurement anchoring the amplitude on large scales, there are
now tight limits as to how much power can be removed on short
scales$^{15,16}$. In simplistic terms, suppressing bubble growth requires that
$H$ decreases as a function of time, and hence horizon-crossing scale. As the
density perturbation spectrum is proportional to $H^2$, one expects
perturbations leaving the horizon at later times to be smaller.

\vspace{6pt}
\vspace{-0.06in}
\begin{center}
{\bf {\it Void Constraints}}
\end{center}

\vspace{6pt}
\vspace{-0.06in}
The heuristic void constraint discussed above has been much improved by Liddle
and Wands$^{17}$, and extended beyond the JBD model$^{14}$. Starting from a
similar expression for the spectrum at the end of inflation, they incorporate
a detailed treatment of the void evolution up to decoupling, conservatively
assuming the voids fill rapidly via relativistic shocks. [This analysis also
depends on the choice of dark matter.] The effect of such a spectrum is then
analysed in detail with regard to various experiments, with the conclusion
that the all-sky coverage provided by COBE$^{18}$ offers the strongest
constraints.

Typically results have to be generated numerically, as in many models even the
void spectrum after inflation cannot be expressed analytically. An example
constraint is in the JBD theory with cold dark matter, which gives
\begin{equation}
\omega_{{\rm {\scriptscriptstyle CDM}}} < 20 + 0.7 \log_{10} \frac{M}{m_{Pl}}
\end{equation}
Although the improvement in $\omega$ appears not too great, this actually
corresponds to a reduction in the volume in large voids by a factor of around
$100$ compared to the earlier constraint. It is also worth emphasising that
even this constraint was derived by neglecting many effects which would in
principle generate anisotropies, conceivably larger ones than were calculated.
Thus we believe that even this void constraint is very conservative, with
exclusions better than 95\%.

\vspace{6pt}
\vspace{-0.06in}
\begin{center}
{\bf {\it Density Perturbations and Gravitational Waves}}
\end{center}

\vspace{6pt}
\vspace{-0.06in}
For convenience and clarity we restrict discussion to the JBD model here.
JBD extended inflation has the convenient property$^{19,20}$ of being
conformally equivalent to the well-investigated power-law inflation model $a
\propto t^p$, where the conformal transformation yields $2p= \omega + 3/2$.
This enables known results to be used directly.

The density perturbation spectrum is tilted from scale-invariance to a
power-law
\begin{equation}
P(k) \propto k^n \quad ; \quad n = 1 - \frac{2}{p-1} = \frac{2\omega - 9}
	{2\omega - 1}
\end{equation}
Power-law inflation generates substantial gravitational waves$^{21,22}$ (also
with a power-law spectrum), and their relative contribution $R$ to large angle
microwave background anisotropies is independent of multipole and given
by$^{23,15}$
\begin{equation}
R = \frac{\Sigma_l^2 ({\rm grav})}{\Sigma_l^2 ({\rm scalar})} = \frac{12.4}{p}
	= \frac{50}{2\omega+3}
\end{equation}
where $\Sigma_l^2$ is the expectation of the square of the $l$-th multipole.
For the allowed $\omega$ values, gravitational waves are the dominant
contributors.

We obtain constraints by utilising the COBE $10^0$ result$^{17}$ with error
bars doubled to give something like a $2$-sigma result: $\sigma_{10^0}^2 =
(1.1 \pm 0.4) \times 10^{-5}$. We include the gravitational wave contribution.
The spectral slopes and amplitudes are uniquely determined by the parameters
$M$ and $\omega$.

\vspace{12pt}
\begin{center}
{\bf COMBINED CONSTRAINTS}
\end{center}

\vspace{12pt}
\vspace{-0.16in}
The constraints on the inflation parameters $M$ and $\omega$ are plotted in
figure 1. All lines are 95\% exclusions or better. The microwave anisotropy
line applies regardless of the choice of dark matter. The void constraint line
does depend on this choice as it governs the efficiency of void filling, and
we plot both cold and hot dark matter results to indicate the spread. One
immediately sees that regardless of $M$, $\omega$ is constrained to be less
than about $17$, and hence $n$ can be no larger than 0.76. The inflaton mass
scale is also strongly constrained.


\begin{figure}
\vspace*{-1.3in}
\special{insert eifig1.dat}

\vspace*{5.4in}
\begin{center}
{\bf Figure 1}
\end{center}
\vspace*{-0.3in}
\end{figure}

Can such a tilt in the spectrum, coupled with the dominant gravitational
contribution to the COBE result, be compatible with small-scale measurements?
In these models the answer is a clear no. Figure 2 considers the specific
choice of standard CDM, and plots constraints on the amplitude and tilt of the
perturbation spectrum. The COBE range implies a very low small-scale amplitude
($\sigma_{8,{\rm {\scriptscriptstyle CDM}}}$ is the variance of the mass in
$8h^{-1}$ Mpc spheres, often indicated as the inverse of the bias parameter)
at small $n$. One can compare this with several experiments. As an example, we
choose constraints on the amplitude from the QDOT survey$^{24}$, following a
procedure of Efstathiou, Bond and White$^{25}$, which indicate quite a high
short-scale amplitude in order to explain peculiar velocities. Recent
results$^{26}$ from comparison of POTENT with IRAS are similar. One can see
immediately that the region $n< 0.84$ is excluded. In fact, things are
probably much worse even than that, because more recent microwave
results$^{27}$ make it extremely unlikely that the true value is towards the
top of the COBE range. And one can see that if the central COBE value is
instead taken as a limit, the constraint tightens by a considerable amount.

\begin{figure}
\vspace*{-1.3in}
\special{insert eifig2.dat}

\vspace*{5.4in}
\begin{center}
{\bf Figure 2}
\end{center}
\vspace*{-0.3in}
\end{figure}

With an admixture of hot dark matter$^{28}$ or a choice $\Omega<1$ (with a
cosmological constant to retain spatial flatness)$^{25}$ things again get
worse, because yet more short-scale power is removed (remember our
normalisation to COBE is dark matter independent). The slight weakening of the
void constraint is unable to compensate for this worsening state of affairs.
One can thus say that {\em any} model of extended inflation which shares the
dynamics of the JBD theory during inflation is convincingly ruled out.

\vspace{12pt}
\begin{center}
{\bf CONCLUSIONS}
\end{center}

\vspace{-0.25in}
\begin{itemize}
\item The combination of big bubble and perturbation constraints is severe.
\item The suppression of big bubbles is inextricably linked to the breaking of
scale-invar\-iance of the perturbation spectra.
\item Models sharing the JBD dynamics are excluded. In addition it is easy to
show the Barrow--Maeda $\omega(\Phi)$ model$^{12}$, which has a more stringent
bubble constraint than the JBD model, is also excluded$^{14}$.
\end{itemize}

To our knowledge, this rules out all extended inflation models bar two.
\begin{enumerate}
\item The Steinhardt--Accetta hyperextended inflation model$^{13}$, utilising
a means of ending inflation dubbed `mode 2' by Crittenden and
Steinhardt$^{29}$. In this model, inflation ends not by bubble collisions but
instead by the complex dynamics ending inflation while the bulk of the
universe is still hung up in false vacuum. The phase transition later
completes in the post-inflationary universe. There are no significant large
bubbles and hence no bubble constraint. However, obtaining working models is
tricky$^{14}$ and probably requires the introduction of a mass for the
gravitational scalar.
\item The `plausible' double inflation model$^{11}$. This relies on a
subsequent phase of slow-roll inflation following the original first-order
inflation, in order to erase some of the effects of the bubbles. This can
weaken the constraints sufficiently to allow a working model.
\end{enumerate}
It is well worth remarking that neither of these models end inflation at the
bubble collision phase. It is fair to say then that there appears no known
working model in which inflation ends via bubble collisions.

To end, we should mention the escape clause. While the bubble constraints
always apply, the perturbation constraints assume that the inflationary
perturbations are responsible for large scale structure. If one can suppress
these further and invoke another means of forming structure, then they can be
evaded. One such option would be to form cosmic strings after or near the end
of inflation. However, the constraint on $M$ implies a reheat temperature of
no more than a few times $10^{14}$ GeV. Perhaps this can be achieved with
global strings.

\vspace{8pt}
\begin{center}
{\bf ACKNOWLEDGEMENTS}
\end{center}

\vspace{8pt}
\vspace{-0.12in}
\noindent
ARL expresses special thanks to David Wands for the collaborations during
which some of the results used here were obtained, and acknowledges support
from the SERC and use of the Starlink computer at Sussex.

\frenchspacing

\vspace{12pt}
\footnotesize
\begin{center}
{\bf REFERENCES}
\end{center}

\vspace{12pt}
\vspace{-0.22in}
\noindent
\begin{tabbing}
\baselineskip 0.4cm
{\ns 1.~~~~~~~}\={\ns L}A, {\ns D. \& P. J. S}TEINHARDT. {\ns 1989. Phys. Rev.
	Lett. {\bf 62:} 376.}\\
{\ns 2.} \> {\ns K}OLB, {\ns E. W.} {\ns 1991. Physica Scripta {\bf T36:}
	199.}\\
{\ns 3.} \> {\ns A}DAMS, {\ns F. C., K. F}REESE, {\ns \& A. H. G}UTH. {\ns
	1990. Phys. Rev. {\bf D43:} 965.}\\
{\ns 4.} \> {\ns L}IDDLE, {\ns A. R. \& D. W}ANDS. {\ns 1992a. Phys. Lett.
	{\bf B276:} 18.}\\
{\ns 5.} \> {\ns G}UTH, {\ns A. H.} {\ns 1981. Phys. Rev. {\bf D23:} 347.}\\
{\ns 6.} \> {\ns A}DAMS, {\ns F. C. \& K. F}REESE. {\ns 1991. Phys Rev {\bf
	D43:} 353.}\\
{\ns 7.} \> {\ns W}EINBERG, {\ns E. J.} {\ns 1989. Phys. Rev. {\bf D40:}
	3950.}\\
{\ns 8.} \> {\ns L}A, {\ns D., P. J. S}TEINHARDT, {\ns \& E. B}ERTSCHINGER.
	{\ns 1989. Phys. Lett. {\bf B231:} 231.}\\
{\ns 9.} \> {\ns R}EASENBERG, {\ns R. D.} \underline{et al.} {\ns 1979.
	Astrophys. J. Lett. {\bf 234:} L219.}\\
{\ns 10.} \> {\ns H}OLMAN, {\ns R., E. W. K}OLB, {\ns \& Y. W}ANG. {\ns 1990.
	Phys. Rev. Lett. {\bf 65:} 17.}\\
{\ns 11.} \> {\ns H}OLMAN, {\ns R.} \underline{et al.} {\ns 1991. Phys. Rev.
	{\bf D43:} 3833}\\
{\ns 12.} \> {\ns B}ARROW, {\ns J. D. \& K. M}AEDA. {\ns 1990. Nucl. Phys.
	{\bf B341:} 294.}\\
{\ns 13.} \> {\ns S}TEINHARDT, {\ns P. J. \& F. S. A}CCETTA. {\ns 1990. Phys.
	Rev. Lett. {\bf 64:} 2740.}\\
{\ns 14.} \> {\ns L}IDDLE, {\ns A. R. \& D. W}ANDS. {\ns 1992b. Phys. Rev.
	{\bf D45:} 2665.}\\
{\ns 15.} \> {\ns L}IDDLE, {\ns A. R. \& D. H. L}YTH. {\ns 1992. Phys. Lett.
	{\bf B291:} 391.}\\
{\ns 16.} \> {\ns L}IDDLE, {\ns A. R. \& D. H. L}YTH. {\ns 1993. To appear,
	Phys. Rep.}\\
{\ns 17.} \> {\ns L}IDDLE, {\ns A. R. \& D. W}ANDS. {\ns 1991. Mon. Not. Roy.
	astr. Soc. {\bf 253:} 637.}\\
{\ns 18.} \> {\ns S}MOOT, {\ns G. F.} \underline{et al.} {\ns 1992. Astrophys.
	J. Lett. {\bf 396:} L1.}\\
{\ns 19.} \> {\ns K}OLB, {\ns E. W., D. S. S}ALOPEK, {\ns \& M. S. T}URNER.
	{\ns 1990. Phys. Rev. {\bf D42:} 3925.}\\
{\ns 20.} \> {\ns L}YTH, {\ns D. H. \& E. D. S}TEWART. {\ns 1992. Phys. Lett.
	{\bf B274:} 168.}\\
{\ns 21.} \> {\ns A}BBOTT, {\ns L. F. \& M. B. W}ISE. {\ns 1984. Nucl. Phys.
	{\bf B244:} 541.}\\
{\ns 22.} \> {\ns S}TAROBINSKY, {\ns A. A.} {\ns 1985. Sov. Astron. Lett. {\bf
	11:} 113.}\\
{\ns 23.} \> {\ns F}ABBRI, {\ns R., F. L}UCCHIN, {\ns \& S. M}ATARRESE. {\ns
	1986. Phys. Lett. {\bf B166:} 49.}\\
{\ns 24.} \> {\ns S}AUNDERS, {\ns W., M. W. R}OWAN-{\ns R}OBINSON, {\ns \& A.
	L}AWRENCE. {\ns 1992. Mon.}\\
    \>  {\ns Not. Roy. astr. Soc. {\bf 258:} 134.}\\
{\ns 25.} \> {\ns E}FSTATHIOU, {\ns G., J. R. B}OND, {\ns \& S. D. M. W}HITE.
	{\ns 1992. Mon. Not. Roy.}\\
    \>  {\ns astr. Soc. {\bf 258:} 1p.}\\
{\ns 26.} \> {\ns D}EKEL, {\ns A.} \underline{et al.} {\ns 1992.
	Princeton preprint IASSNS-AST 92/55.}\\
{\ns 27.} \> {\ns G}AIER, {\ns T.} \underline{et al.} {\ns 1992. Astrophys. J.
	Lett. {\bf 398:} L1.}\\
{\ns 28.} \> {\ns D}AVIS, {\ns M., F. J. S}UMMERS, {\ns \& D. S}CHLEGEL. {\ns
	1992. Nature {\bf 359:} 393.}\\
{\ns 29.} \> {\ns C}RITTENDEN, {\ns R. \& P. J. S}TEINHARDT. {\ns 1992. Phys.
	Lett. {\bf B293:} 32.}
\end{tabbing}
\end{document}